\documentclass[aps,twocolumn,floatfix,prb]{revtex4}

\usepackage[pdftex]{graphicx}
\usepackage{epsfig}
\usepackage{amsmath}
\def\notes#1{ } 
\def\be{\begin{equation}}
\def\ee{\end{equation}}
\def\bea{\begin{eqnarray}}
\def\eea{\end{eqnarray}}

\def\bra #1 {\langle {#1} \vert}
\def\ket #1 {\vert {#1} \rangle}

\begin{document}
\renewcommand{\thefootnote}{\fnsymbol{footnote}}
\renewcommand{\theequation}{\arabic{section}.\arabic{equation}}

\title{Stochastic determination of effective Hamiltonian for the full configuration interaction solution of quasi-degenerate electronic states}

\author{Seiichiro Ten-no}
\email[]{E-mail: tenno@cs.kobe-u.ac.jp}

\affiliation{Graduate School of System Informatics, Department of Computational Sciences, Kobe University, Nada-ku, Kobe 657-8501, Japan}

\date{\today}

\begin{abstract}
We propose a novel quantum Monte Carlo method in configuration space, which stochastically samples the contribution from a large secondary space to the effective Hamiltonian in the energy dependent partitioning of L\"owdin.
The method treats quasi-degenerate electronic states on a target energy with bond dissociations and electronic excitations avoiding significant amount of the negative sign problem.
The performance is tested with small model systems of H$_4$ and N$_2$ at various configurations with quasi-degeneracy.
\end{abstract}

\maketitle
\section{INTRODUCTION}
Quantum Monte Carlo (QMC) methods in configuration space\cite{afqmc03, afqmc07, ohtsuka08, ohtsuka10, ohtsuka11, fciqmc09, i-fciqmc} have recently drawn attention of quantum chemists for obtaining accurate correlation energies of {\it ab initio} electronic structure theory.
In these methods, the ground state wave functions are obtained stochastically by a Monte Carlo simulation in the configuration space typically of Slater determinants.
The auxiliary field QMC (AFQMC) method approximately controls the phase problem by writing the imaginary-time propagator of a many-body system with two-body interactions in terms of propagators for independent particles interacting with external AF.\cite{afqmc03, afqmc07}
More naive treatments are the projector Monte Carlo (PMC) method of Ohtsuka and Nagase\cite{ohtsuka08, ohtsuka10, ohtsuka11} and the full configuration interaction QMC (FCIQMC) method of Booth, Alavi and coworkers,\cite{fciqmc09, i-fciqmc} which directly sample the coefficients in a FCI problem using walkers on configuration space basis functions.
A long imaginary-time evolution of the wave function of
\be
\Psi(\tau) =e^{-\tau(\hat H - E)}\Psi(0)
\ee
shrinks the contributions of all states with finite excitation energies exponentially to magnify the ground state solution. 
Then the propagation for a short imaginary-time is calculated by a simulation of the differential equation,
\be
\frac{d\Psi(\tau)}{d\tau} =-(\hat H - E)\Psi(\tau).\label{eq:prop}
\ee
Apart from the capability for applications, the schemes of the population dynamics of walkers in PMC and FCIQMC are similar except for the treatment of the diagonal contribution of Hamiltonian matrix.

The QMC methods suffer from a sign problem which prohibits stable energy convergence.
The initiator variant of FCIQMC\cite{i-fciqmc} ($i$-FCIQMC) overcomes the problem by controlling the growth of walkers of sign-incoherent progeny in an approximate manner with a bias of surviving criteria on a matrix with a sign problem.
It is demonstrated that $i$-FCIQMC reduces the prefactor of exponential scaling significantly,\cite{scale-i-fciqmc} and the method has been successfully applied to the calculations of first-row diatomic molecules\cite{fciqmc-frdim} and even to those of real solids.\cite{fciqmc-solids}
More recently, $i$-FCIQMC has been generalized in such a way that the propagation of a set of important determinants is performed by deterministic projection, resulting in a less severe sign problem and a large reduction in the statistical error.\cite{sqmc}
The basis set truncation error can be reduced by combining the QMC methods with F12 electronic structure theory.\cite{fciqmc-f12, pmc-f12}
A method for obtaining isolated excited states within the FCIQMC framework has been also proposed very recently.\cite{fciqmc-excit}

Despite the effort devoted to the advancements, none of the aforementioned QMC methods is effective for electronic states in a quasi-degenerated or degenerated situation.
This is because those QMC deal with only a single-root problem in the Hilbert space.
In this paper, we propose a novel QMC scheme to calculate the effective Hamiltonian of the L\"owdin partitioning rather than a single-root energy.
The method treats multiple solutions near a target energy in a model space simultanously.
This fact enables us to apply the new QMC method for various interesting phenomena in chemistry and physics involving bond dissociations and electronic excitations.

\section{METHOD}
\subsection{Effective Hamiltonian and imaginary-time evolution of the transfer matrix}
In this section, we formulate the basic properties of effective Hamiltonian and its imaginary-time evolution.
Let us begin with the L\"owdin partitioning technique\cite{Loewdin1, Loewdin2} for the exact effective Hamiltonian.
The entire configuration space is divided into the model space P containing solution(s) of interest and its complement Q.
The Schr\"odingier equation in the partitioned basis is in the form
\be
\begin{pmatrix}
{\bf H}_{\rm PP} & {\bf H}_{\rm PQ} \\
{\bf H}_{\rm QP} &{\bf H}_{\rm QQ}
\end{pmatrix}
\begin{pmatrix}
{\bf C}_{\rm P}\\
{\bf C}_{\rm Q}
\end{pmatrix}
 = E
\begin{pmatrix}
{\bf C}_{\rm P}\\
{\bf C}_{\rm Q}\\
\end{pmatrix}.
\ee
For the state energy $E$, the CI vectors ${\bf C}_{\rm P}$ and ${\bf C}_{\rm Q}$ are explicitly related as
\bea
{\bf C}_{\rm Q} &=& {\bf T}_{\rm QP} {\bf C}_{\rm P}, \label{eq:l-amp}\\
{\bf T}_{\rm QP} &=& -({\bf H}_{\rm QQ} - E {\bf I}_{\rm QQ})^{-1} {\bf H}_{\rm QP}, \label{eq:l-condition}
\eea
where ${\bf T}_{\rm QP}$ is the transfer matrix, and (\ref{eq:l-condition}) is called the energy dependent partitioning (EDP) condition.
${\bf C}_{\rm Q}$ is then eliminated from the Schr\"odinger equation as
\be
{\bf H}^{\rm eff}_{\rm PP} {\bf C}_{\rm P} = E {\bf C}_{\rm P},
\ee
with the effective Hamiltonian,
\bea
{\bf H}^{\rm eff}_{\rm PP} &=& {\bf H}_{\rm PP} + {\boldsymbol \Sigma}_{\rm PP}, \label{eq:effh1}\\
{\boldsymbol \Sigma}_{\rm PP} &=& {\bf H}_{\rm PQ} {\bf T}_{\rm QP}. \label{eq:effh2}
\eea
According to the expansion of the model space, the coupling between P- and Q-spaces through the resolvent in (\ref{eq:l-condition}) is diminished systematically as the gap between $E$ and the eigen values of ${\bf H}_{\rm QQ}$ increases.
The EDP technique is applicable to any excited state by changing the target energy $E$ as long as the principal component of the state is involved in the model space.
In this case, ${\bf C}_{\rm P}$ for different $E$ are non-orthogonal except for the perfect degeneracy.

One notes that an effective Hamiltonian appointing all solutions in the P-space universally is determined by the eigenvalue independent partitioning (EIP) with the same number of variables,\cite{EIP1, EIP2}
\be
{\bf H}_{\rm QP} + {\bf H}_{\rm QQ} {\bf T}_{\rm QP} - {\bf T}_{\rm QP}({\bf H}_{\rm PP} + {\bf H}_{\rm PQ} {\bf T}_{\rm QP})=0.
\ee
We do not take this strategy in this paper because the numerical  treatment of the linear equation (\ref{eq:l-condition}) is more straightforward for the purpose of QMC implementation.
The solution of the EIP equation is not unique due to the nonlinear structure of the form.
If all of the states in the model space are well-separated from those in the Q-space, the transfer matrix ${\bf T}_{\rm QP}$ can be expanded in a Rayleigh-Schr\"odinger perturbation series,\cite{RS-Lindgren} which often accompanies intruder state problems.

The imaginary-time evolution (\ref{eq:prop}) consists of P- and Q-space contributions, $\Psi=\Psi_{\rm P}+\Psi_{\rm Q}$.
The evolution of $\Psi_{\rm Q}$ is obtained by multiplying the projector onto the Q-space, $\hat Q=1-\hat P$,
\be
\frac{d\Psi_{\rm Q}(\tau)}{d\tau} =-\hat Q[( \hat H - E )\Psi_{\rm Q}(\tau)+ \hat H \Psi_{\rm P}(\tau)],
\ee
or in the CI coefficients,
\be
\frac {d{\bf C}_{\rm Q}}{d\tau}=- ({\bf H}_{\rm QQ}-E{\bf I}_{\rm QQ}){\bf C}_{\rm Q}(\tau)-{\bf H}_{\rm QP}{\bf C}_{\rm P}(\tau).
\ee
Substituting (\ref{eq:l-amp}) into the above, and breaking up the summation over ${\bf C}_{\rm P}(\tau)$, we obtain the propagation of the transfer matrix,
\be
\frac{d{\bf T}_{\rm QP}(\tau)}{d\tau}= -({\bf H}_{\rm QQ} - E {\bf I}_{\rm QQ}){\bf T}_{\rm QP}(\tau) - {\bf H}_{\rm QP}.\label{eq:tel}
\ee
The stationary solution of the equation $d{\bf T}_{\rm QP}(\tau)/d\tau=0$ satisfies the EDP condition (\ref{eq:l-condition}).
Then the exact effective Hamiltonian ${\bf H}^{\rm eff}_{\rm PP}$ is obtained from (\ref{eq:effh1}) and (\ref{eq:effh2}).
For a given state energy, the ${\bf T}_{\rm QP}(\tau)$ propagation is decoupled for each P-space determinant, and the $N_P$ equations can be treated independently.
Eq. (\ref{eq:tel}) bears close resemblance with the coefficients form of (\ref{eq:prop}) except for the additional contribution of $-{\bf H}_{\rm QP}$.
This fact enables us to develop a QMC algorithm in a manner similar to the other QMC methods in configuration space as explained below.

\subsection{Quantum Monte Carlo algorithm}
Henceforth, we use the notations $I,J,...$ and $A,B,...$ for the Slater determinants in the P- and Q-spaces, respectively.
The dimension of the secondary space Q can be prohibitively large.
We therefore intend to treat the imaginary-time evolution of ${\bf T}_{\rm QP}(\tau)$ stochastically with a small storage for ${\bf H}^{\rm eff}_{\rm PP}$ in the model space QMC algorithm (MSQMC).
For this purpose, we augment (\ref{eq:tel}) by multiplying a diagonal matrix ${\bf N}^{\rm (b)}_{\rm PP}$ containing fixed values of booster weights for the P-space determinants, and propagate it with a small time interval $\delta\tau$,
\be
\delta {\bf T'}_{\rm QP}(\tau)= \delta \tau [{\boldsymbol \rho}_{\rm QP}^{\rm (D)}(\tau) + {\boldsymbol \rho}_{\rm QP}^{\rm (Q)}(\tau) + {\boldsymbol \rho}_{\rm QP}^{\rm (P)}(\tau)],
\ee
where ${\boldsymbol \rho}_{\rm QP}^{\rm (D)}$, ${\boldsymbol \rho}_{\rm QP}^{\rm (Q)}$, and ${\boldsymbol \rho}_{\rm QP}^{\rm (P)}$ are diagonal (D), off-diagonal Q-space (Q), and P-space (P) contributions, respectively,
\bea
{\boldsymbol \rho}_{\rm QP}^{\rm (D)}(\tau) &=& -({\bf H}^{\rm (D)}_{\rm QQ} - E {\bf I}_{\rm QQ}){\bf T'}_{\rm QP}(\tau), \label{eq:D} \\
{\boldsymbol \rho}_{\rm QP}^{\rm (Q)}(\tau) &=& -{\bf H}^{\rm (O)}_{\rm QQ}{\bf T'}_{\rm QP}(\tau), \label{eq:Q} \\
{\boldsymbol \rho}_{\rm QP}^{\rm (P)}(\tau) &=& -{\bf H}_{\rm QP}(\tau){\bf N}^{\rm (b)}_{\rm PP}, \label{eq:P}
\eea
${\bf H}^{\rm (D)}_{\rm QQ}$ and ${\bf H}^{\rm (O)}_{\rm QQ}$ are diagonal and off-diagonal matrices of ${\bf H}_{\rm QQ}$, and ${\bf T'}_{\rm QP}(\tau)={\bf T}_{\rm QP}(\tau){\bf N}^{\rm (b)}_{\rm PP}$.
Although the booster weights can be fractional, we assume them positive integers for simplicity in this work.
The MSQMC method for the effective Hamiltonian represents ${\bf T'}_{\rm QP}(\tau)$ as collections of walkers,
\be
T'_{AI}(\tau) = \sum\limits_{\mu \in I} {s_\mu \delta_{\kappa_\mu A}}. \label{eq:instT}
\ee
$s_\mu$ and $\kappa_\mu$ are the sign and corresponding determinant of the walker $\mu$, respectively, and the distribution of walkers is dependent on $\tau$.
The Monte Carlo steps corresponding to (\ref{eq:D}), (\ref{eq:Q}), and (\ref{eq:P}) are independent for each walker as well as the P-space determinant $I$.
Thus the merged walkers of different imaginary time steps represent the solution of linear equation with much smaller number of walkers compared to the true dimension of ${\bf T}_{\rm QP}(\tau)$ as
\be
T'_{AI} = \frac{1}{\Delta \tau} \int_{\tau_{\rm i}}^{\tau_{\rm f}} {d\tau \sum\limits_{\mu \in I}{s_\mu \delta_{\kappa_\mu A}}},
\ee
where $\tau_{\rm i}$ and $\tau_{\rm f}$ are the initial and final times for the imaginary time integration, and $\Delta \tau=\tau_{\rm f}-\tau_{\rm i}$ is the interval.
For a fixed $N^{\rm (b)}_{I}$, the number of walkers $N^{\rm (w)}_{I}$ fluctuates during the imaginary-time evolution.
The average number of walkers $\bar N^{\rm (w)}_{I}$ is dependent on the coupling of $I$ and the Q-space. 
Instantaneous $T_{AI}$ is obtained by removing the booster weight, $T_{AI}=T'_{AI}/N^{\rm (b)}_{I}$.
We then accumulate ${\bf H}^{\rm eff}_{\rm PP}$ with the instantaneous ${\bf T}_{\rm QP}$.
In what follows is the practical algorithm of MSQMC, in which the individual steps of death/cloning, spawning, and annihilation are entirely parallel to those in FCIQMC.\cite{fciqmc09}
\begin{enumerate}
\item Secular part: Diagonalize ${\bf H}^{\rm eff}_{\rm PP}$ for ${\bf C}_{\rm P}$ and $E$.
${\bf H}^{\rm eff}_{\rm PP}={\bf H}_{\rm PP}$ for the first iteration.

\item QMC part:
All operations are reflected in a new list of walkers, and there is no interference until the annihilation step.
\begin{enumerate}
\item diagonal death/cloning step of ${\boldsymbol \rho}_{\rm QP}^{\rm (D)}(\tau)$: Each walker in the Q-space dies with the probability
\be
p_\mu^{\rm (D)}=\delta \tau (H_{\kappa_\mu \kappa_\mu}-E).
\ee
Although the cloning event for $p_\mu^{\rm (D)}<0$ does not happen with a normal choice of $E$, a walker may be cloned with the probability $|p_\mu^{\rm (D)}|$.
The treatment of $|p_\mu|>1$ follows the way in the  FCIQMC algorithm.\cite{fciqmc09} 

\item Q-space spawning step of ${\boldsymbol \rho}_{\rm QP}^{\rm (Q)}(\tau)$: For each walker $\mu$ in the Q-space, a candidate is chosen randomly from the interacting determinants with the probability $1/n_{\kappa_\mu}$, where $n_\kappa$ is the number of Slater determinants in the Q-space interacting with $\kappa$ excluding $\kappa$ itself.
Then a new walker $\nu$ is spawned with the probability
\be
p_\nu^{\rm (Q)}=-\delta \tau n_{\kappa_\mu} H_{\lambda_\nu \kappa_\mu}s_\mu,
\ee
for $p_\nu^{\rm (Q)}>0$.
If $p_\nu^{\rm (Q)}<0$, a walker with opposite sign is spawned with probability $|p_\nu^{\rm (Q)}|$.
\item P-space spawning step of ${\boldsymbol \rho}_{\rm QP}^{\rm (P)}(\tau)$: Inside a loop over a discretized index of the booster weight for each $I$, a determinant is chosen randomly as in the Q-space spawning, and a new walker in the Q-space $\nu$ is spawned with the probability
\be
p_\nu^{\rm (P)}=-\delta \tau n_{I} H_{\kappa_\nu I},
\ee
for $p_\nu^{\rm (P)}>0$, where $n_I$ is the number of Slater determinants in the Q-space interacting with $I$.
If $p_\nu^{\rm (P)}<0$, a walker with opposite sign is spawned with probability $|p_\nu^{\rm (P)}|$.
\item Annihilation step: Remove pair determinants with opposite signs from the list of walkers.
\item Sigma step: Using the list of new walkers, accumulate ${\boldsymbol \Sigma}_{\rm PP}$ as
\be
\Sigma_{IJ} = \frac{1}{\Delta\tau N^{\rm (b)}_{J}}\int_{\tau_{\rm i}}^{\tau_{\rm f}}d\tau \sum\limits_{\mu \in J} {s_\mu H_{I\kappa_\mu}}.
\ee
for effective Hamiltonian ${\bf H}^{\rm eff}_{\rm PP}$ in the symmetric form of (\ref{eq:effh1}).
\end{enumerate}
For a probability $p>1$, the event is caused int$(p)$ times immediately with an additional event with the probability $p-$int$(p)$.\cite{fciqmc09}
The steps (a) - (e) are repeated until the limit of the micro cycle.
\end{enumerate}
The secular and QMC parts are repeated during the macro iteration cycle.

\subsection{Stochastic promotion/demotion stage}
For partitioning of the spaces in the MSQMC method, the P-space is preferred to be compact to shrink the computational cost.
Contrarily, the presence of high-wighted amplitudes on determinants in the Q-space deteriorates the accuracy of MSQMC.
In the static mode of MSQMC, the partitioning can be performed by selecting predominating determinants from a complete active space (CAS) CI, which is generally unavailable for large molecules.

For more general applications, we introduce a stochastic promotion/demotion (SPD) stage for the model space exploiting a sampling of QMC on top of a long  production run.
After certain QMC steps, SPD moves a determinant in the Q-space to the P-space if the number of walkers on the determinant exceeds $T_{\rm p} N^{\rm (t)}$, where $T_{\rm p}$ is a promotion threshold, and $N^{\rm (t)}=N^{\rm (b)}+N^{\rm (w)}$ is the sum of the booster weight and number of walkers.
Nevertheless, the stochastic promotion can choose unwanted Slater determinants with small amplitudes.
Thus, the determinants with amplitudes smaller than a threshold, $|C_I| < T_{\rm d}$, are demoted in the subsequent secular step.

\section{RESULTS AND DISCUSSIONS}
We examine the performance of MSQMC on the model systems of H$_4$ and N$_2$ molecules, in which a full diagonalization is possible with a Hamiltonian matrix on memory.
SPD is performed with the contracted basis of the solution of ${\bf H}_{\rm PP}$ rather than the individual Slater determinants using the first macro iteration.
The lowest 10 Slater determinants in $H_{II}$ are employed for the initial ${\bf H}_{\rm PP}$ of SPD.
An $N^{\rm (t)}$ constant simulation is performed during SPD.
Population control is accomplished by the simple scaling $N^{\rm (b)}_{\rm new}=N^{\rm (b)}N^{\rm (t)}/(N^{\rm (b)}+N^{\rm (w)})$ every 10 steps.
We use the default parameters of the promotion/demotion thresholds, $T_{\rm p}=1\times10^{-3}$ and $T_{\rm d}=1 \times 10^{-2}$.
The energy obtained from a diagonalization of  ${\bf H}_{\rm PP}$ is usually sufficiently low, and no population control is employed for MSQMC after SPD in this work.
The total number of walkers is temporarily increased to $N^{\rm (t)}=2\times10^4$ for SPD, while the booster weight is kept $N^{\rm (b)}=5\times10^2$ for all $I$ in MSQMC.
The death/cloning step is associated with the instantaneous energy until the integration is turned on at the 5th macro iteration.
The definition of the instantaneous energy in this work is the expectation value of the instantaneous contribution of the distribution of walkers Eq. (\ref{eq:instT}) to the effective Hamiltonian using CI coefficients fixed in each macro iteration.
Each macro iteration contains $10^{3}$ steps with $\delta \tau = 0.01$.
Three simulations using different seeds for random numbers are performed for each method to confirm the magnitude of statistical error.
We do not employ any bias in the sampling space of walkers to avoid introducing additional error albeit the initiator extension of MSQMC is straightforward

\begin{figure}[t]
\begin{center}
\includegraphics[width=250pt]{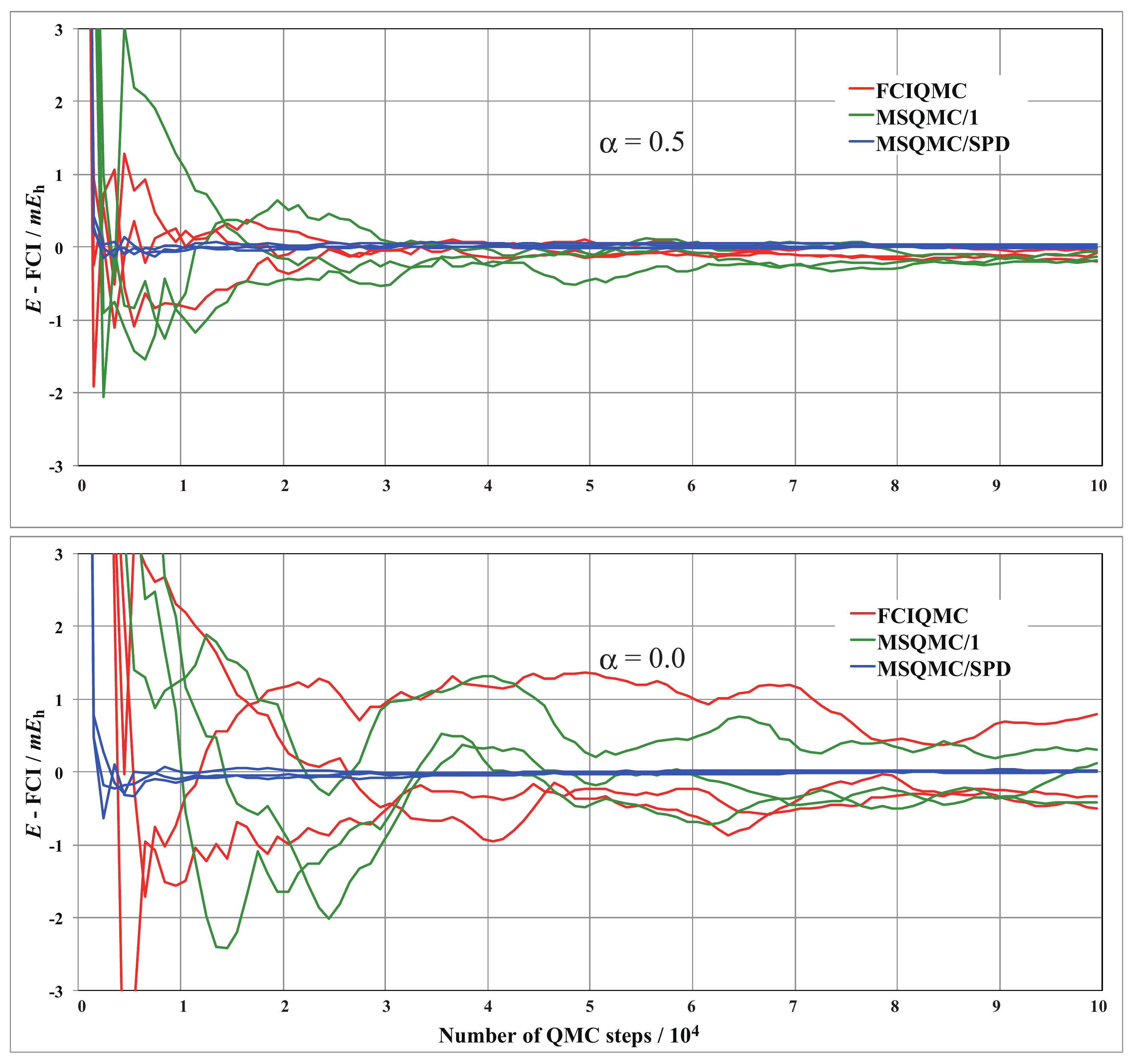}
\end{center}
\caption{Behavior of the deviation of $E$ from FCI for the H$_4$ model. Three simulations with different seeds are performed for each method and geometry. The SPD stage promoted 44 and 50 Slater determinants for $\alpha=0.0$ and 0.5 respectively.}
\label{fig:h4_1}
\end{figure}

\begin{table}
\caption
{\label{tab:h4_sd}
Statistical measures ($mE_{\rm h}$) of all energies at every macro iteration cycle after $5\times10^{4}$ QMC steps.}
\begin{tabular}{lcccc}
\hline
& FCIQMC & \multicolumn{3}{c}{MSQMC} \\
\cline{3-5}
& & $N_{\rm P}=1$ & $N_{\rm P}=2$ & SPD \\
\hline
$\alpha=0.5$ \\
$\bar\Delta_{\rm abs}$& 0.09 & 0.19 & $-$ & 0.02 \\
$\bar\Delta_{\rm max}$& 0.19 & 0.49 & $-$ & 0.06 \\
$\Delta_{\rm std}$& 0.06 & 0.12 & $-$ & 0.02 \\
$\Delta_{\rm std}^{\rm (s)}$\footnote[1]{Standard deviation of energies with different seeds at the 50th macro iteration step.}& 0.12 & 0.22 & $-$ & 0.03 \\
\\
$\alpha=0.0$ \\
$\bar\Delta_{\rm abs}$& 0.53 & 0.36 & 0.14 & 0.01 \\
$\bar\Delta_{\rm max}$& 1.31 & 0.77 & 0.26 & 0.03 \\
$\Delta_{\rm std}$& 0.62 & 0.37 & 0.15 & 0.02 \\
$\Delta_{\rm std}^{\rm (s)}$$^a$& 0.97 & 0.38 & 0.22 & 0.01 \\
\hline
\end{tabular}
\end{table}

\begin{figure}[t]
\begin{center}
\includegraphics[width=250pt]{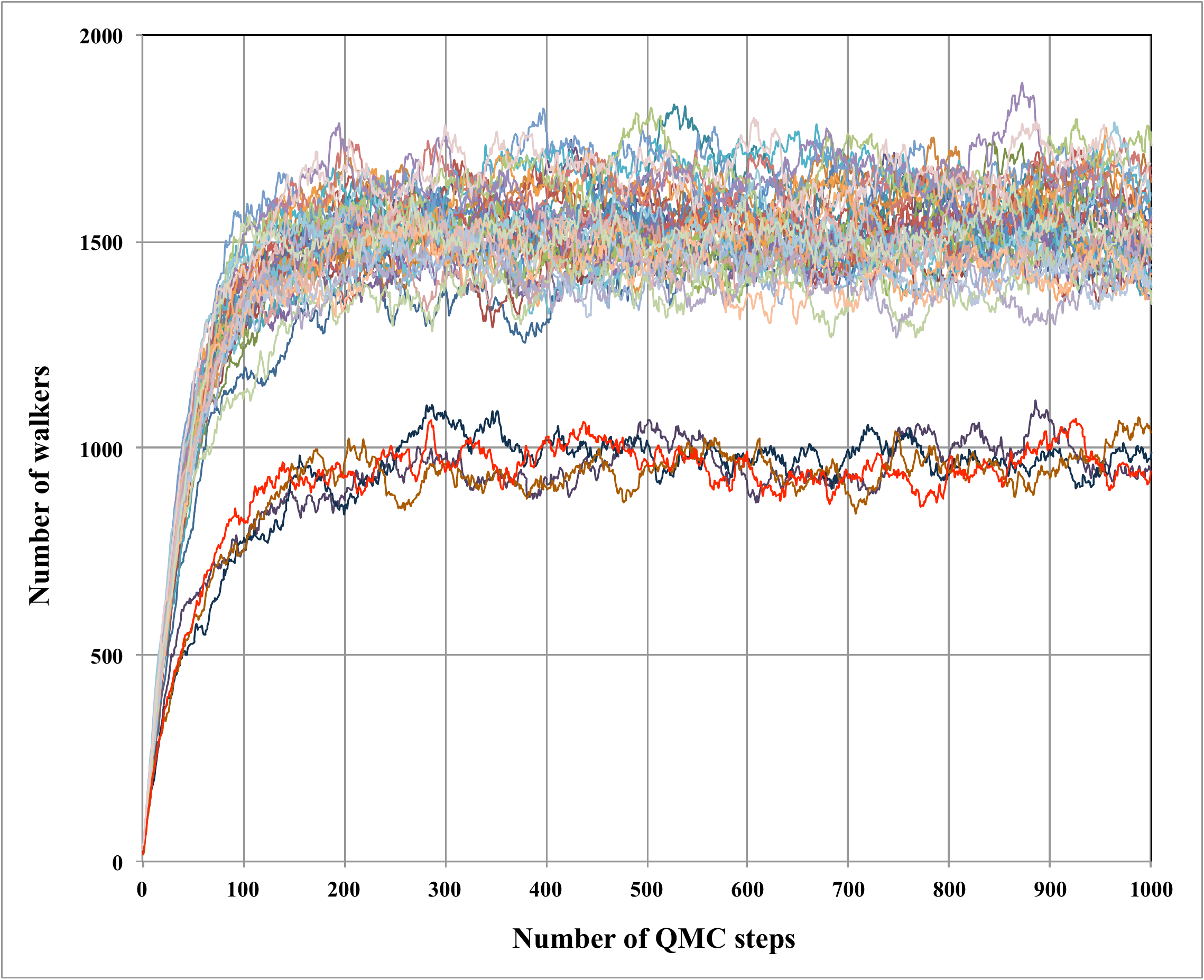}
\end{center}
\caption{Growth of the numbers of walkers in the first 1,000 steps of the MSQMC/SPD simulation with 44 promoted Slater determinants in the model space for $\alpha=0.0$.}
\label{fig:walkers}
\end{figure}

\begin{figure}[t]
\begin{center}
\includegraphics[width=250pt]{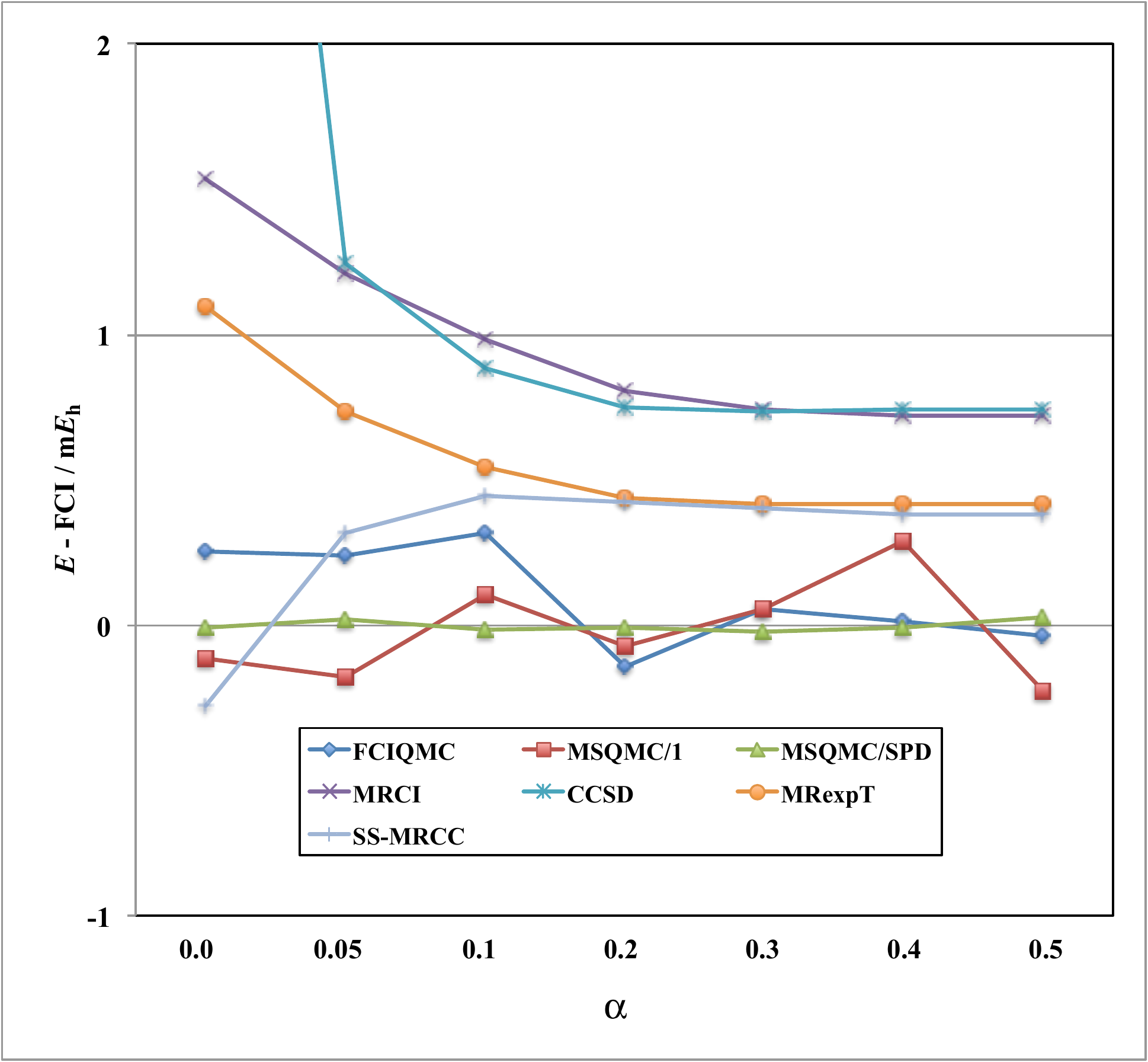}
\end{center}
\caption{Errors of QMC and correlated methods for different $\alpha$ of the H$_4$ model. The results for the deterministic methods are taken from Refs. \cite{Uttam99, Uttam11, Mike06}.}
\label{fig:h4}
\end{figure}

\subsection{H$_4$ model}
The first example is the H$_4$ model with the double zeta plus polarization basis set.
This model has been employed frequently to test multi-reference electronic structure methods.\cite{Uttam99, Uttam11, Mike06}
The non-dynamic correlation effects are significant near the regular square form with the angle parameter $\alpha=0$.
The dimensions of the FCI problems are 5,068 and 5,524 for $\alpha=0$ and 0.5, respectively, in $D_{2h}$ point group symmetry, and 10,036 for other $\alpha$ in $C_{2v}$ symmetry.
Fig. \ref{fig:h4_1} shows the behaviors of errors in $E$ compared to FCI with respect to the number of QMC steps for $\alpha=0.5$ and $0.0$.
The number of walkers grows to nearly 4,000 for $\alpha=0$ in the MSQMC simulation only with the Hartree-Fock (HF) determinant in the P-space (MSQMC/1).
Correspondingly, we set the initial number of walkers $N^{\rm (w)}=4,000$  in a FCIQMC simulation with the logarithmic control of energy shift.\cite{fciqmc09}
A perfect $N^{\rm (w)}$ constant simulation is not possible with the poor initial HF energy, and $N^{\rm (w)}$ finally grows by ca. 20\% .
For FCIQMC, the integrated value of local energy with intermediate normalization is used for comparison of the results.
For $\alpha=0.5$, the HF determinant is predominating, and all results are accurate to 0.5 $mE_{\rm h}$ after $3 \times 10^4$ steps.
FCIQMC compares with MSQMC only with the HF determinant in the P-space denoted by MSQMC/1 while MSQMC/SPD outperforms them significantly.
Table \ref{tab:h4_sd} shows statistical measures of energy, mean absolute error $\bar\Delta_{\rm abs}$, maximum absolute error $\bar\Delta_{\rm max}$, and standard deviation $\Delta_{\rm std}$.
$\Delta_{\rm std}$ is in agreement with $\bar\Delta_{\rm abs}$ in each simulation.
$\Delta_{\rm std}^{\rm (s)}$ of energies with different seeds at the 50th macro iteration step also compares with $\Delta_{\rm std}$ as a good candidate for the measure of error estimation.
$\alpha=0.0$ is the most difficult geometry for H$_4$.
Clearly the accuracies of FCIQMC and MSQMC/1 are deteriorated compared to the results for $\alpha=0.5$.
In this geometry, $(a_g)^2 (b_{2u})^2$ and $(a_g)^2 (b_{3u})^2$ determinants in the $D_{2h}$ have equal weights.
The extended model space with the 2 Slater determinants, MSQMC/2, clearly improves the result.
MSQMC/SPD is as accurate as the result for $\alpha=0.5$.
Based on $\bar\Delta_{\rm abs}$ and $\Delta_{\rm std}$, MSQMC/SPD is 1-2 orders of magnitude more  accurate than FCIQMC or MSQMC/1 especially for the quasi-degenerate situation.

Fig. \ref{fig:walkers} shows the growing behavior of the numbers of walkers in MSQMC/SPD for $\alpha=0$.
Each line corresponds to the number for one of the Slater determinants in the model space promoted at the SPD stage.
The number distribution grows very rapidly in the initial 200 steps.
There are two groups in the equilibrated distribution around $N^{\rm (w)}=1,000$ and 1,500, indicating the presence of the main component of the wave function and determinants interacting relatively strongly with those in the Q-space, respectively.
In either case, the numbers of walkers needed for the description of the Q-space contribution are noticeably reduced on average from 4,000 in the MSQMC/1 case.
The accuracy of MSQMC/QMC is supported by the reduced coupling between the P- and Q-spaces.

It is tempting to compare the accuracy of the QMC methods with those of correlated methods available in {\it ab initio} electronic structure theory.
Fig. \ref{fig:h4} compare results for various $\alpha$ parameters.
The averaged energy after $5 \times 10^4$ steps of simulations with different seeds is used as the resulting value of a scheme for all QMC methods.
Compared to CCSD, all multi-reference methods supplies well-balanced dependency of the geometrical parameter $\alpha$.
The errors of FCIQMC and MSQMC/1 are smaller than those of all deterministic methods on average.
Nevertheless, these QMC methods will not be extremely useful for practical applications due to the unsystematic errors.
MSQMC/SPD significantly reduces the error; the result is accurate to a few 10 $\mu E_{\rm h}$ for all $\alpha$.

\begin{widetext}
\begin{center}
\begin{table}
\caption
{\label{tab:n2}
Ground and excited state energies of the N$_2$ molecule at several bond distances.}
\begin{tabular}{lrrrrrrrr}
\hline
& \multicolumn{2}{c}{$1A_g$} && \multicolumn{2}{c}{$2A_g$} && $2A_g-1A_g$ &\\
\cline{2-3} \cline{5-6} \cline{8-9}
& $E/E_{\rm h}$ & ($\Delta_{\rm std}^{\rm (s)}$) && $E/E_{\rm h}$ & ($\Delta_{\rm std}^{\rm (s)}$) && $\Delta E$/eV \\
\hline
$r=2.068$ $a_0$ \\
FCI& -109.142 767 &&& -108.414 054 &&& 19.829\\
FCIQMC & 0.000 135 & (0.000 185) &&-&&&-\\
MSQMC/1 & 0.000 287 & (0.000 318) &&-&&&-\\
MSQMC/SPD/$E_1$ & 0.000 022 & (0.000 013) && 0.016 824 & (0.000 020) && 0.457 \\
MSQMC/SPD/$E_2$ & -0.007 198 & (0.000 161) && -0.000 030 & (0.000 096) && 0.195 \\
MSQMC/SPD/$E_2-E_1$\footnote[1]{Excitation energy as the difference of the target energies of MSQMC/SPD/$E_1$ and $E_2$ simulations.} &&&&&&& -0.001\\
\hline
$r=4.2$ $a_0$ \\
FCI& -108.807 006 &&& -108.791 832 &&& 0.413\\
FCIQMC & 0.003 768 & (0.003 758) &&-&&&-\\
MSQMC/1\footnote[2]{The reduced number of virtual walkers $N^{\rm (b)}=100$ is used to keep $N^{\rm (t)}$ comparable.} & -0.001 082 & (0.005 063) &&-&&&-\\
MSQMC/SPD/$E_1$ & 0.000 083 & (0.000 059) && 0.002 117 & (0.001 760) && 0.055 \\
MSQMC/SPD/$E_2$ & -0.000 699 & (0.000 430) && 0.000 237 & (0.000 281) && 0.025 \\
MSQMC/SPD/$E_2-E_1$$^a$ &&&&&&& 0.004\\
\hline
$r=6.0$ $a_0$ \\
FCI& -108.801 215 &&& -108.800 615 &&& 0.016 \\
FCIQMC & 0.001 026 & (0.004 316) &&-&&&- \\
MSQMC/1$^b$ & -0.006 170 & (0.005 178) &&-&&&- \\
MSQMC/SPD/$E_1$ & -0.000 076 & (0.000 142) && -0.000 073 & (0.000 163) && 0.000 \\
MSQMC/SPD/$E_2$ & -0.000 121 & (0.000 210) && 0.000 114 & (0.000 127) && 0.006 \\
MSQMC/SPD/$E_2-E_1$$^a$ &&&&&&& 0.005\\
\hline
\end{tabular}
\end{table}
\end{center}
\end{widetext}
\subsection{Bond stretching with electronic excitation of N$_2$}
The second system is the N$_2$ molecule at different bond distances.
Despite the small dimension of the FCI problem (8,152 in $D_{2h}$ symmetry), it is not easy for QMC to to deal with the system since several CI coefficients with equally large amplitudes are present at a stretched distance.
We use the same parameters as the H$_4$ case except for the MSQMC/1 simulation, in which the weight of HF determinant is insignificant at large $r$ to increase $N^{\rm (w)}$.
We therefore use the reduced value of $N^{\rm (b)}=100$ at $r=4.2$ and $6.0$ $a_0$ to make the number of walkers comparable with those in FCIQMC.
Such a significant growth does not occur in MSQMC/SPD.
The FCI space is generated by CASSCF calculations with 10 electrons in 10 orbitals (10,10) with the aug-cc-pVDZ basis set.\cite{Bas_CC1,Bas_CC2}
The orbitals are in the supplementary material.\cite{supmat}

The results for the ground and excited sate energies are summarized in Table \ref{tab:n2}.
Although the exact solutions of the two lowlying states in the $A_g$ symmetry are uncoupled $^1\Sigma_g^+$ and $^5\Sigma_g^+$, they cannot be separated by symmetry in the basis of Slater determinants.
The number of the model space determinants generated in the SPD stage ranges from 50 to 80 dependent on the distance. 
Near the equilibrium geometry, $r=2.068$ $a_0$, the ground state is dominated by a single Slater determinant, and FCIQMC and MSQMC/1 reveal accurate.
We have performed energy specific MSQMC simulations appointing the ground and excited states separately as denoted by MSQMC/SPD/$E_1$ and $E_2$.
The energy of the target state in each simulation is calculated precisely in spite of the fact that the SPD stage is performed for the ground state.
The effective Hamiltonian is not universally accurate with somewhat large excitation energies due to the large energy gap between the states at $r=2.068$ $a_0$.
The difference between the two energy specific results supplies the excitation energy very close to the FCI value.

At the extended bond distance $4.2$ $a_0$, the energy gap is reduced to 0.4 eV.
The quasi-degeneracy clearly makes FCIQMC and MSQMC/1 imprecise.
$\Delta_{\rm std}^{\rm (s)}$ of the methods amount to several $mE_{\rm h}$ indicating a significant negative-sign problem induced from the neighboring states.
The results of MSQMC/SPD retains similar accuracies for the target states.
The reduction of the excitation energy increases the transferability of the effective Hamiltonian with the entire errors less than 0.1 eV.
These features hold for the results at $6.0$ $a_0$.
In this case, the energy gap is comparable to the static error of the MSQMC/SPD simulations.

\subsection{Perspectives on real applications}

The examination of the preliminary implementation of MSQMC motivated by the newly developed formula for the time evolution of the transfer matrix (\ref{eq:tel}) has clearly demonstrated the effectiveness of the method for electronic states of small model systems with quasi-degeneracy and excitations.
So the remaining question  is, whether the effectiveness is preserved on scaling up to realistic systems with large CI dimensions of interest in chemistry and physics?
Indeed, MSQMC, which only cures the instability from quasi-degeneracy by including interacting states into ${\bf H}^{\rm (eff)}$, is not a sort of means to eliminate an exponential growth in the number of necessary walkers due to sign-incoherent spawning events from determinants with low populations for larger systems or larger basis sets.
The initiator method introduced by the Alavi group\cite{i-fciqmc, scale-i-fciqmc} has ameliorated the situation in FCIQMC reducing the sign problem, which smoothly disappears as increasing the number of walkers. 
Likewise, it is expected that the initiator approach will complement MSQMC achieving the effectiveness for real systems.
More specifically, the initiator space at first equivalent to the P-space of MSQMC is enlarged such that a noninitiator determinant in the Q-space with a walker number exceeding a threshold is made an initiator eligible to contribute to the spawning processes onto unoccupied determinants.
In this way, the initiator method will be incorporated into MSQMC to produce an extremely useful scheme capable of treating large and difficult systems with quasi-degeneracy and strong electron correlation.
The initiator extension is meaningful for larger systems.
A comparison of the initiator variants, {\it i}-FCIQMC and {\it i}-MSQMC is therefore of interest and should be made with full-scale implementations.

\section{conclusions}
In this paper, we have proposed and tested the MSQMC method, which incorporates with the effective Hamiltonian of the L\"owdin partitioning technique into a stochastic simulation for the FCI contribution outside the model space.
The new method complements most of the defects in the previous QMC methods in configuration space.
The main features of MSQMC are summarized in the following.
\begin{enumerate}
\item The most severe negative-sign problem is originating from a stochastic treatment of the competition of a small number of predominating Slater determinants.
MSQMC sidesteps the problem by handling the principal component in a deterministic secular equation with an effective Hamiltonian in the same spirit as the semi-stochastic QMC method.\cite{sqmc}
\item MSQMC is capable of treating multi-root solutions in the vicinity of the target energy.
This feature guarantees accurate descriptions of degenerate and quasi-degenerate electronic states involving a bond-breaking.
\item The EDP condition holds for excited states with different target energies.
MSQMC is capable of providing an accurate energy of excited state with a model space containing basis functions predominating the state of interest.
The effective Hamiltonian is transferable for states in a certain range of the target energy.
\item The MSQMC steps are independent for each parental determinant in the model space.
This fact enables an efficient parallel implementation by communicating the state energy and effective Hamiltonian in each macro iteration, in addition to the intrinsic scalability with respect to the number of processes with different seeds for random numbers as well as the numbers of walkers in each of them.
\item The stochastic promotion/demotion step is particularly effective to extract important basis functions to describe the target state.
\end{enumerate}
MSQMC is computationally demanding compared to the standard electronic structure methods and other QMC methods for a single-root solution.
Nevertheless, the requirement for a FCI problem shifts from the storage and communication of CI coefficients to computing resources, which can be easily obtained in modern computational environments.
An implementation with on-the-fly computation of Hamiltonian matrix elements for large-scale FCI problems will be reported elsewhere.\cite{OT}\

\begin{acknowledgements}
The author is indebted to Yuhki Ohtsuka for valuable discussions and providing the FCI Hamiltonian matrices tested in this work.
He also thanks Debashis Mukherjee for suggestions on the principles of effective Hamiltonian of EIP, Cyrus Umrigar for information about the semi-stochastic QMC method and useful comments on the manuscript, and George Booth, Ali Alavi, and Masatoshi Imada for precious communications about the paper.
This work is partly supported by the Grant-in-Aids for Scientific Research (B) (No. 00270471) from the Japan Society for the Promotion of Science (JSPS), and the Strategic Programs for Innovative Research (SPIRE), MEXT, and the Computational Materials Science Initiative (CMSI), Japan.
\end{acknowledgements}

\end{document}